\documentstyle[epsfig]{mn2e}
\newif\ifAMStwofonts
\AMStwofontstrue

\def\be{\begin{equation}}
\def\ee{\end{equation}}

\def\Msun{{M_\odot}}
\def\Zsun{{Z_\odot}}

\def\gsim{\lower.5ex\hbox{\gtsima}}
\def\lsim{\lower.5ex\hbox{\ltsima}}
\def\gtsima{$\; \buildrel > \over \sim \;$}
\def\ltsima{$\; \buildrel < \over \sim \;$}
\def\prosima{$\; \buildrel \propto \over \sim \;$}
\def\gsim{\lower.5ex\hbox{\gtsima}}
\def\lsim{\lower.5ex\hbox{\ltsima}}
\def\simgt{\lower.5ex\hbox{\gtsima}}
\def\simlt{\lower.5ex\hbox{\ltsima}}
\def\simpr{\lower.5ex\hbox{\prosima}}
\def\ie{{\frenchspacing\it i.e. }}

\title[Pop III stars: hidden or disappeared?]{Population III stars: hidden or disappeared ?}

\author[Luca Tornatore, Andrea Ferrara \& Raffaella Schneider]{Luca Tornatore$^1$, Andrea Ferrara$^1$ \& Raffaella Schneider$^{2}$\\
  $1$ SISSA/Internation School for Advanced Studies, Via Beirut 4, 34100 Trieste, Italy\\
  $2$ INAF/Osservatorio Astrofisico di Arcetri, Largo Enrico Fermi 5,
  50125 Firenze, Italy\\}

\date{Feb 01, 2007}

\pagerange{\pageref{firstpage}--\pageref{lastpage}}
\pubyear{2007}

\begin{document}

\maketitle
\label{firstpage}

\begin{abstract}
  A Pop~III/Pop II transition from massive to normal stars is
  predicted to occur when the metallicity of the star forming gas
  crosses the critical range $Z_{cr}=10^{-5\pm 1} Z_\odot$. To
  investigate the cosmic implications of such process we use 
  numerical simulations which follow the evolution, metal
  enrichment and energy deposition of both Pop III and Pop II stars.
  We find that: (i) due to inefficient heavy element transport by
  outflows and slow "genetic" transmission during hierarchical growth,
  large fluctuations around the average metallicity arise; as a result
  Pop III star formation continues down to $z=2.5$, but at a low peak
  rate of $10^{-5} M_\odot {\rm yr}^{-1} {\rm Mpc}^{-3}$ occurring at
  $z\approx 6$ (about $10^{-4}$ of the Pop~II one); (ii) Pop III
  star formation proceeds in a "inside-out" mode in which formation
  sites are progressively confined at the periphery of collapsed
  structures, where the low gas density and correspondingly long
  free-fall timescales result in a very inefficient astration. These
  conclusions strongly encourage deep searches for pristine star
  formation sites at moderate ($2<z<5$) redshifts where metal free
  stars are likely to be hidden.
\end{abstract}


\section{Introduction}

The physical conditions in primordial star-forming regions appear to
systematically favor the formation of very massive stars. This is due
to the combined effect of the larger gas fragmentation scale and
accretion rate, and the very limited opacity. On the other hand,
observations of present-day stellar populations (Pop II/I stars) show
that stars form according to a Salpeter Initial Mass Function (IMF)
with a characteristic mass of $\approx 1 \Msun$, below which the IMF flattens.
Thus, unless the current picture of primordial star formation is
lacking in some fundamental ingredient, a transition between these two
modes of star formation must have occurred at some time during cosmic
evolution.

What are the physical drivers of such transition ? Since the early
study by Yoshii \& Sabano (1980), gas metallicity has been suspected
to play a key role. This idea has been later on expanded and
substantiated by a number of detailed studies (Omukai 2000; Bromm et
al. 2001; Schneider et al. 2002, 2003). The emerging physical
interpretation states that the fragmentation properties of the
collapsing clouds change as the mean metallicity of the gas increases
above a critical threshold, $Z_{\rm cr} = 10^{-5 \pm 1} \; \Zsun$. The
characteristic masses of protostellar gas clouds with $Z<Z_{\rm cr}$ are 
predicted to be relatively large ($>100\;\Msun$), whereas in clouds with
$Z>Z_{\rm cr}$ lower characteristic masses can be formed. Within the critical
metallicity range, low-mass gas clouds can form if a sufficient amount of
metals are depleted onto dust grains, which provide an additional
efficient cooling channel at high density (Schneider et al. 2003;
2006a; Omukai et al. 2005; Omukai \& Tsuribe 2006). According to this
view, the formation of Pop III stars (defined as those with $Z <
Z_{cr}$) is regulated by the rate at which heavy elements are produced
and mixed in the gas surrounding the first star-forming regions ({\it
  chemical feedback}).

In principle, Pop III stars can continue to form until late epochs,
provided that gas pockets of sufficiently low metallicity can be
preserved during cosmic evolution. This condition can be met for newly
formed halos that either (i) gain their gas from regions not yet
polluted by outflows from nearby star forming galaxies, or (ii) have
progenitors in which star formation has not occurred or was suppressed
(Ciardi \& Ferrara 2005). Thus, chemical feedback can act via two
physically different channels, involving either the transport of
metals by outflows or in a "genetic" form, \ie inheriting metals from
the parent sub-halos (Schneider et al. 2006b).

Scannapieco, Schneider \& Ferrara (2003, SSF) studied the relative
importance of the outflow channel, finding that metal transport is
generally inefficient; as a consequence the transition epoch is
extended in time, coeval Pop III and Pop II star formation occurs, and
Pop III stars continue to form down to $z \lsim 5$. Similar
conclusions are reached by Furlanetto \& Loeb (2005). 
Analytic models (Mackey, Bromm \& Hernquist 2003) and high-resolution 
numerical simulations (Yoshida, Bromm \& Hernquist 2004) show that if
most of the early generation stars die as pair-instability supernovae,
the {\it volume-averaged} IGM metallicity will quickly 
reach $Z=10^{-4} Z_\odot$ by $z\approx 15-20$.
However, as shown by SSF and confirmed here, this condition does not
guarantee a self-termination of massive Pop III star formation, due to the
highly inhomogeneous metal distribution.

Additional complications come from the effects of radiative feedback 
(Ricotti, Gnedin \& Shull 2002; Machacek, Bryan, Abel 2003; Omukai \& Yoshii 2003; 
Yoshida et al. 2003; Susa \& Umemura 2006), HD chemistry (Nagakura \& Omukai 2005; Greif \& Bromm
2006; Johnson \& Bromm 2006; Yoshida et al. 2007) and radiative transfer (Ciardi, Ferrara
\& Abel 2000; Ricotti, Gnedin \& Shull 2001; Kitayama et al. 2001).
These are generally found to be important for mini-halos
($T_{vir} < 10^4$~K), or under physical conditions not relevant for this study.

Although the above studies find difficult to rapidly suppress the
formation of Pop III stars, this common wisdom has to face the fact
that no metal-free stars have yet been found by surveys of metal-poor 
stars of the Milky Way halo (Cayrel et al. 2004; Beers \& Christlieb 2005;
Tumlinson, Venkatesan, Shull 2004; Tumlinson 2006; Daigne et al. 2006;
Salvadori, Schneider \& Ferrara 2007). Alternative probes as (i) the 
apparent excess in the cosmic near-infrared background 
(Salvaterra \& Ferrara 2003; Santos, Bromm \& Kamionkowski 2002; 
Salvaterra et al. 2006; Kashlinsky et al. 2006a,b),
(ii) the equivalent width distributions of high-$z$ Ly$\alpha$
emitters (Malhotra \& Rhoads 2002; Dawson et al. 2004), (iii) the
He~II 1640\AA~line in composite spectra of LBGs (Shapley et al. 2003;
Nagao et al. 2005) are yielding only tentative evidence for the
presence of Pop III stars at $z < 9$.

The question then remains: are Pop III stars hidden (due to their very
small statistical frequency or because they reside in yet unexplored
environments) or did they disappear as a result of chemical feedback a
long time ago ? The aim of this Letter is an attempt to address this
question.

\begin{figure}
\psfig{figure=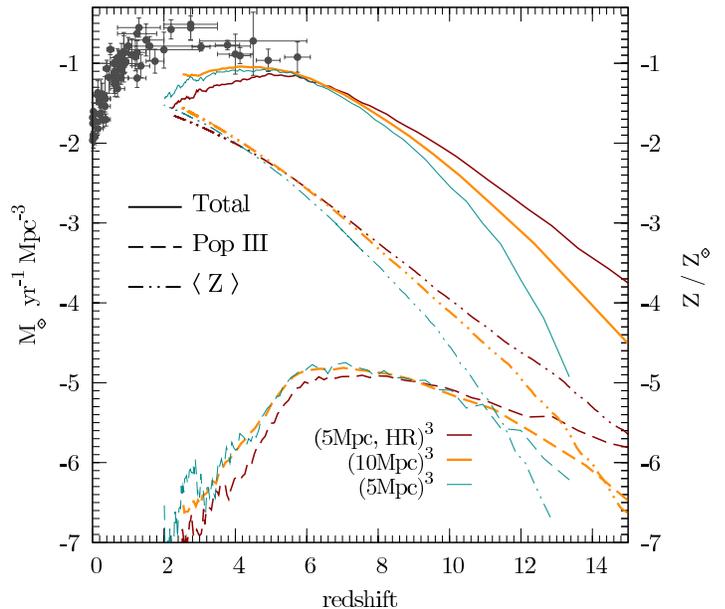,height=8cm}
\caption{Predicted evolution of Pop II (solid lines) and Pop III
  (dashed) cosmic star formation rates, and mass-averaged metallicity
  (dot-dot-dashed). The results of the three different simulation runs
  described in the text are shown for each quantity. As a reference,
  low-redshift measurements (points) taken from Hopkins (2004) are
  reported. }
\vspace{-0.5cm}
\label{sfr}
\end{figure}

\section{Numerical simulations}

For the present study we have performed a set of
cosmological\footnote{Throughout the paper, we adopt a $\Lambda$CDM
  cosmological model with parameters $\Omega_M = 0.26$,
  $\Omega_{\Lambda} = 0.74$, $h=0.73$, $\Omega_b=0.041$, $n=1$ and
  $\sigma_8=0.8$, in agreement with the 3-yr WMAP results (Spergel et
  al. 2006).} simulations using the publicly available code
GADGET\footnote{www.mpa-garching.mpg.de/galform/gadget/} (Springel
2005) with an improved treatment of chemical enrichment as in
Tornatore et al. (2007). For the present study, we have further
implemented the possibility to assign different Initial Mass Functions
(IMF) to each star forming particle, depending on the gas metallicity.
In particular, for the purpose of this work, if $Z < Z_{\rm cr}$, the
adopted IMF is a Salpeter law with lower (upper) limit of $100
M_\odot$ ($500 M_\odot$); only stars in the pair-instability ($140
M_\odot < M < 260 M_\odot$) range contribute to metal enrichment
(Heger \& Woosley 2002). If $Z \ge Z_{\rm cr}$, we assume that the
above limits are shifted to $0.1 M_\odot$ ($100 M_\odot$),
respectively; stars above $40 M_\odot$ end their lives as black holes
swallowing their metals. In the following, we show results for $Z_{\rm
  cr} = 10^{-4}Z_\odot$. These two populations, to which we will refer
to as Pop III and Pop II stars respectively, differ also for their
metal yield (i.e. the fraction of stellar mass transformed into
metals), $y$, and explosion energy per unit mass of baryons going into
stars, $\epsilon$. For Pop III stars we follow Heger \& Woosley (2002)
and assume $y=0.183$, $\epsilon=3.5\times 10^{16}$~erg~g$^{-1}$; for
Pop II stars we adopt a metallicity-dependent value of $y$ taken from
Woosley \& Weaver (1995) and $\epsilon=3.4\times
10^{15}$~erg~g$^{-1}$. Consistently with the above yields, we follow
the production and transport of six different metal species, namely:
C, O, Mg, Si, S, Fe. In our simulations, the IMF depends on the
  value of the gas metallicity. Physically, the gas can be enriched by
  metals released by local stars or through winds powered by stars in
  nearby regions. Thus, the IMF is determined according to the
  SPH-smoothed value\footnote{The Z-smoothing is calculated
    normalizing the kernel in a sub-region of the sph volume (here
    $0.2$ of the sph length), independently of the number of neighbors
    found.} rather than on the intrinsic particle metallicity. For
the purpose of this work, we have chosen to simulate a (comoving)
volume of $L=10 h^{-1}$~Mpc with $N_p=2\times 256^3$ (dark+baryonic)
particles, corresponding to a dark matter (baryonic) particle mass of
$M_p=3.62 \times 10^6 h^{-1} M_\odot$ ($6.83 \times 10^5 h^{-1}
M_\odot$); the corresponding force resolution is $2~$kpc. Our
resolution does not allow to track the formation of mini-halos, whose
stellar contribution remains very uncertain due to radiative feedback
effects (Haiman \& Bryan 2006; Susa \& Umemura 2006; Ahn \& Shapiro
2007).

The computation is initialized at $z=99$ and carried on until
$z=2.5$. In order to check the convergence of the results, we have run
two additional simulations with $L=5 h^{-1}$~Mpc and $N_p=2\times
(256^3,128^3)$, the latter one having the same particle mass as in the
reference run. 

The gas photo-ionization and heating rates are calculated at
equilibrium with a background ionizing radiation due to the combined
contribution of galaxies and quasars, taken from Haardt \& Madau
(1996), shifted so that the intensity at 1 Ryd is $J_\nu= 0.3\times
10^{-21}$~erg ~s$^{-1}$~Hz$^{-1}$, in agreement with Bolton et al.
(2005). Gas cools according to the cooling function given by
Sutherland \& Dopita (1993), corrected for both helium and hydrogen
photoionization; because metal photoionization is not followed
explicitly, the cooling might be somewhat overestimated. Supernova
winds are treated as in the original model by Springel \& Hernquist
(2003); however, for simplicity and because the mass load and kinetic
energy fraction are unknown parameters, we have given the winds from
both population an initial velocity $v_w=500$~km~s$^{-1}$, which
appears to be consistent with that derived from observations of
high-$z$ starburst galaxies (Adelberger et al. 2003; Shapley et al. 2003).  
Wind particles are
  temporarily hydrodynamically decoupled until either (i) they have
  moved by a traveling length $\lambda=2$~kpc, or (ii) their density
  has decreased below 0.5 times the star formation density threshold
  ($n_\star = 0.1$~cm$^{-3}$). The metals are donated by
  star particles to the surrounding gas ones using a SPH kernel as
  described in Tornatore et al. (2007).

\section{Results}

Our simulation outputs contain a large number of useful information
concerning the mode (Pop III or Pop II) of star formation and the
source of metals associated to each particle. In addition, we can
recover if heavy elements have been produced {\it in situ} or
transported to that location by outflows.

We start the analysis of the simulation outputs by reconstructing the
evolution of Pop II and Pop III star formation rates (SFR) and the
corresponding evolution of the mass-averaged metallicity, $\langle Z
\rangle$, due to the dispersal of heavy elements produced by these
stars. These are shown by the curves in Fig. \ref{sfr}. Star formation
begins very early ($z \simgt 15$) for both populations. At these high
redshifts, though, there are considerable uncertainties in the rates
due to numerical resolution effects, as gathered from a comparison
among the three runs reported in Fig. \ref{sfr}. In general,
decreasing the resolution leads to an underestimate of the SFR at high
redshift; smaller volumes, instead, miss the activity at later (and
more easily observable) epochs. Note, however, that the scatter in
the PopIII/PopII SFR ratio remains $<20-30\%$ independent of resolution. 
For this reason we consider $L=10 h^{-1}$~Mpc, $N_p=2\times 256^3$ 
as our fiducial run.

The gas mass enriched to $Z>Z_{\rm cr}$ by a single Pop III star
is so large that Pop II stars are always the dominant star formation
mode (apart the very first event):
at $z=14$ only about 1\% of the stars in the universe are born in formation
sites where $Z<Z_{\rm cr}$, and hence producing Pop III stars. Such ratio
steadily decreases reaching $\approx 10^{-4}$ at $z = 6$ and
rapidly dropping afterwards.

We pause to outline two important points. First, and in agreement with
previous findings by SSF and Schneider et al. 2006, Pop III stars
continue to form well beyond $z=10$, the epoch at which $\langle Z
\rangle > Z_{\rm cr}$, at non-negligible rates: for example, at $z=5$, the
SFR integrated over the corresponding Hubble time would yield a Pop
III stellar density of $\Omega_{III}\approx 2\times 10^{-6} \Omega_b$.
This does not come as a surprise if we note that, due to the highly
inhomogeneous nature of metal enrichment (see below), relatively
pristine regions survive for several Gyr; these are the host
environment for Pop III formation sites. The second point to note is
that, in addition to chemical feedback, the suppression of Pop III
star formation is also caused by the IGM photo-heating due to
reionization, which in our simulation occurs at $z\approx 7$. In fact, the
associated increase of the Jeans (or more precisely, filtering) scale
inhibits the collapse of low mass halos which are more likely to be
relatively uncontaminated (Schneider et al. 2006) with respect to
larger ones which are genetically polluted by their merging
progenitors. 

\begin{figure*}
\centerline{
\psfig{figure=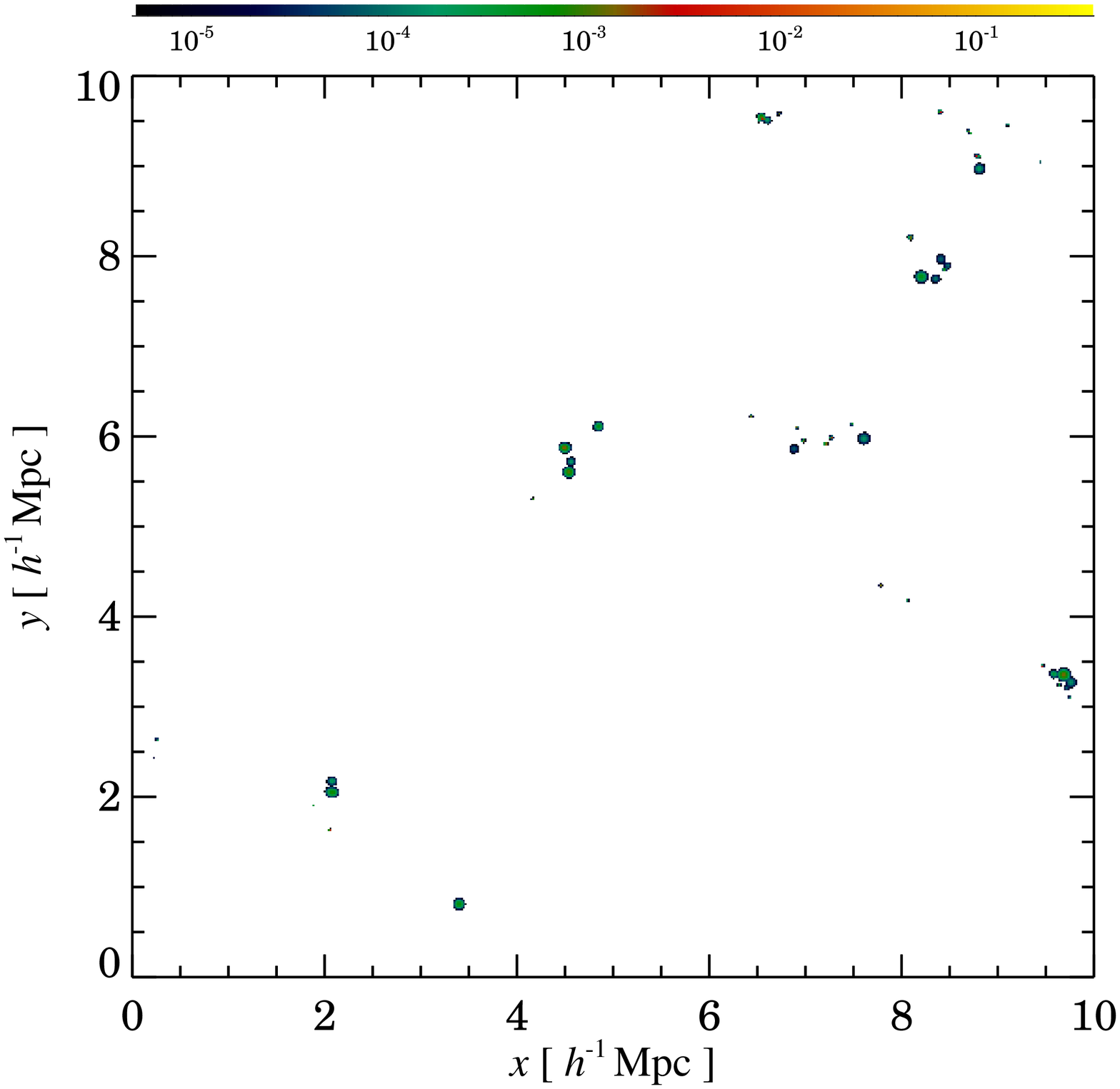,width=8cm,angle=0}
\psfig{figure=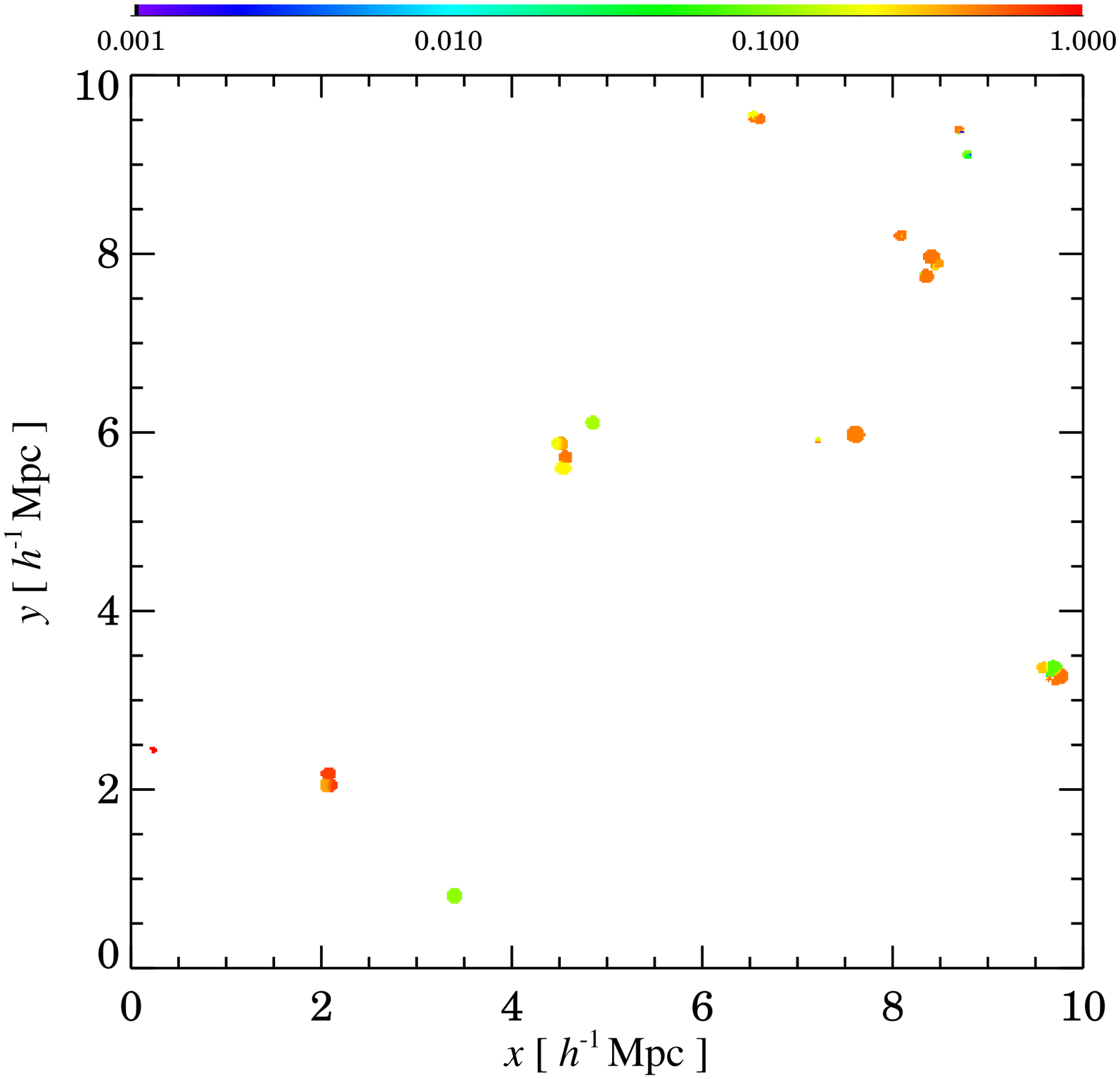,width=8cm,angle=0}}
\vspace{-0.5cm}
\centerline{
\psfig{figure=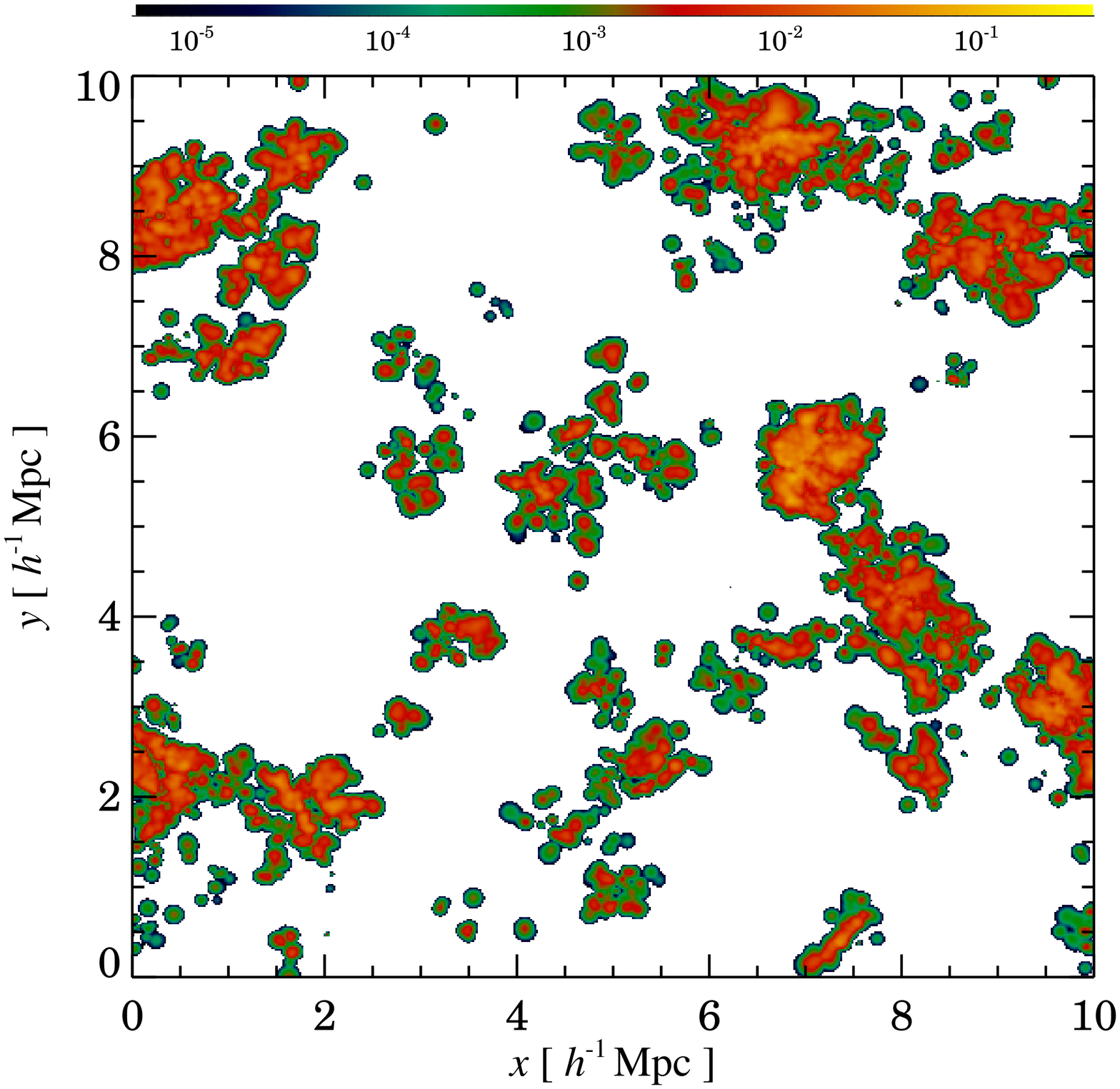,width=8cm,angle=0}
\psfig{figure=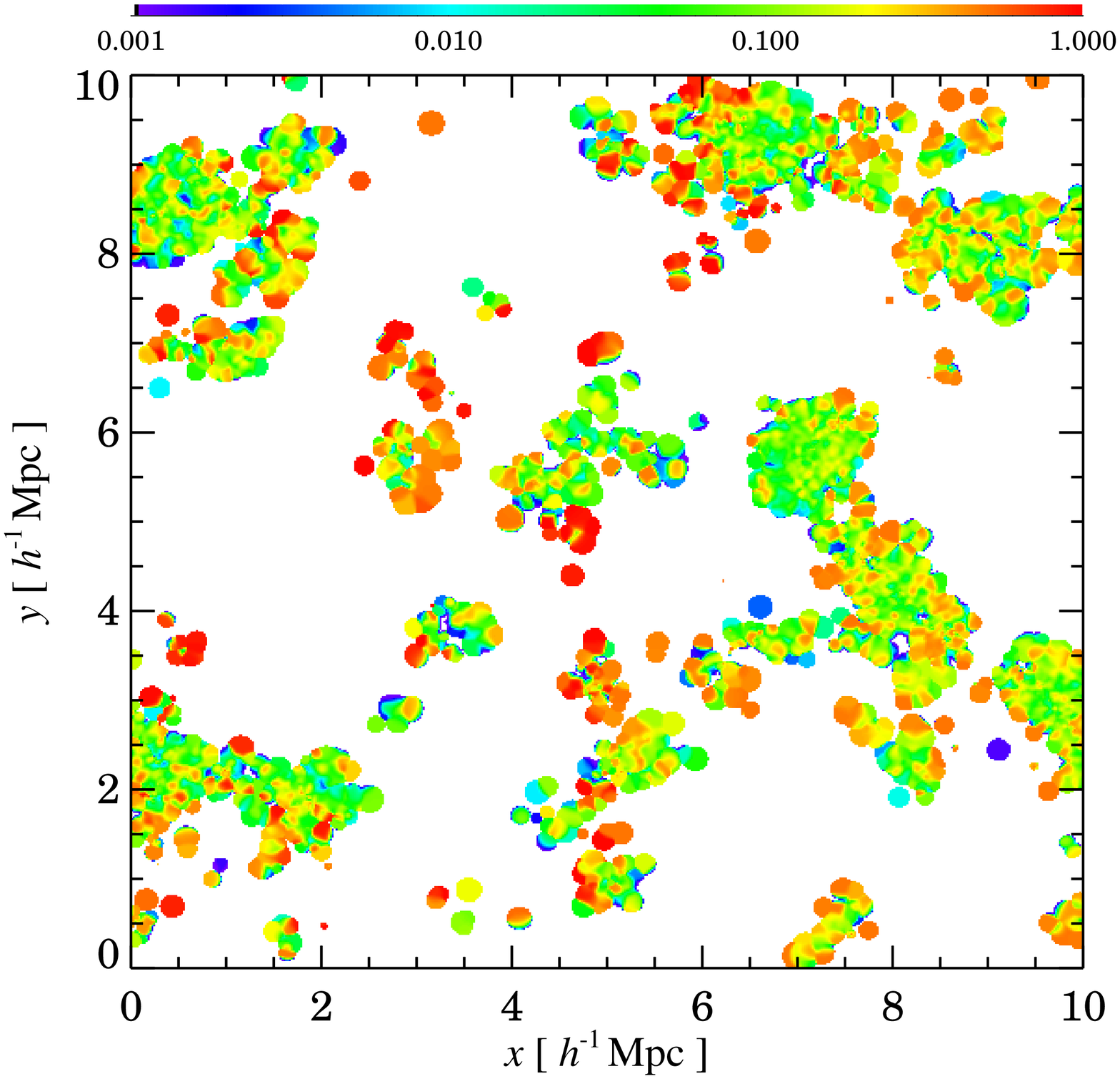,width=8cm,angle=0}}
\vspace{-0.5cm}
\centerline{
\psfig{figure=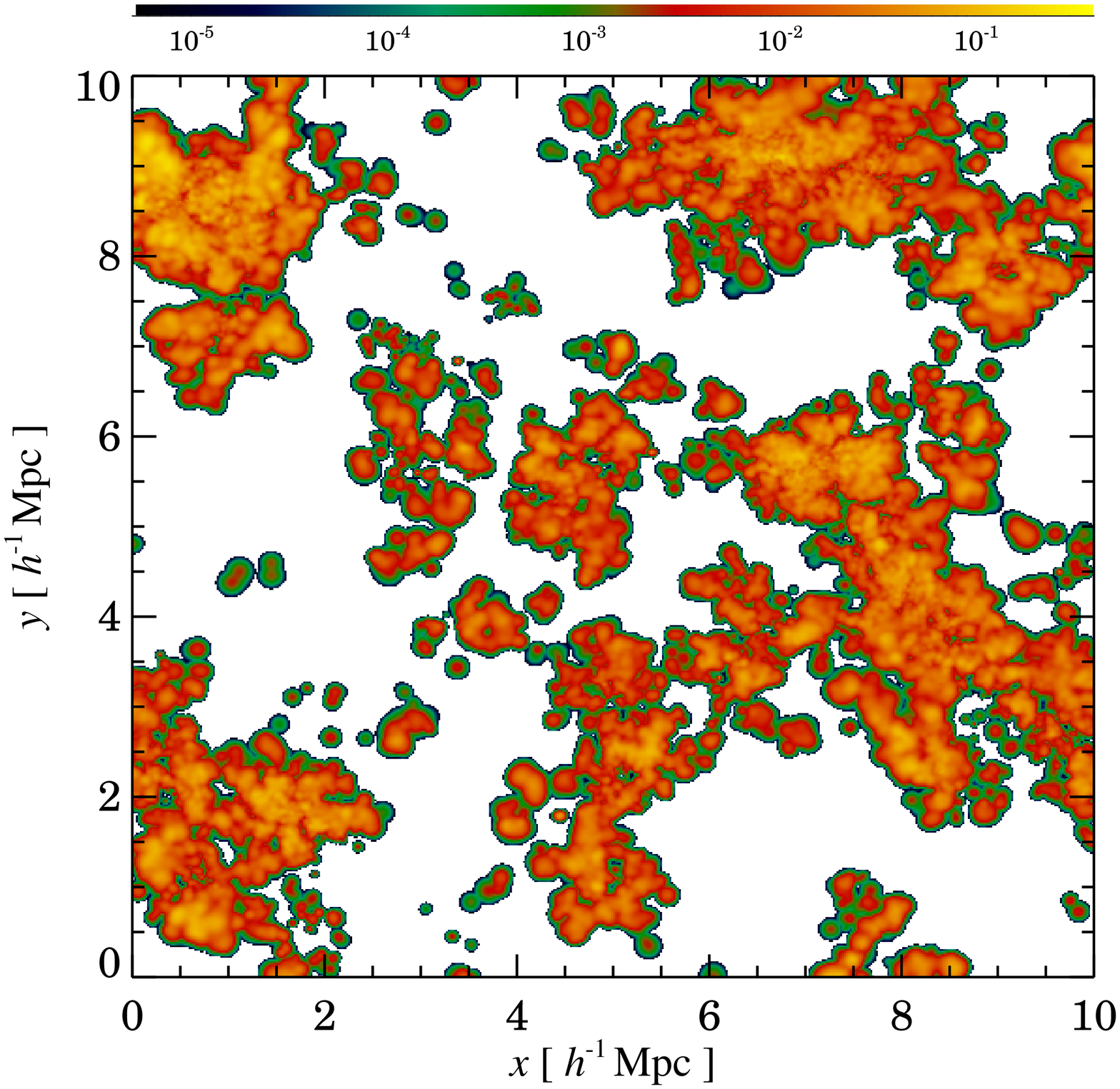,width=8cm,angle=0}
\psfig{figure=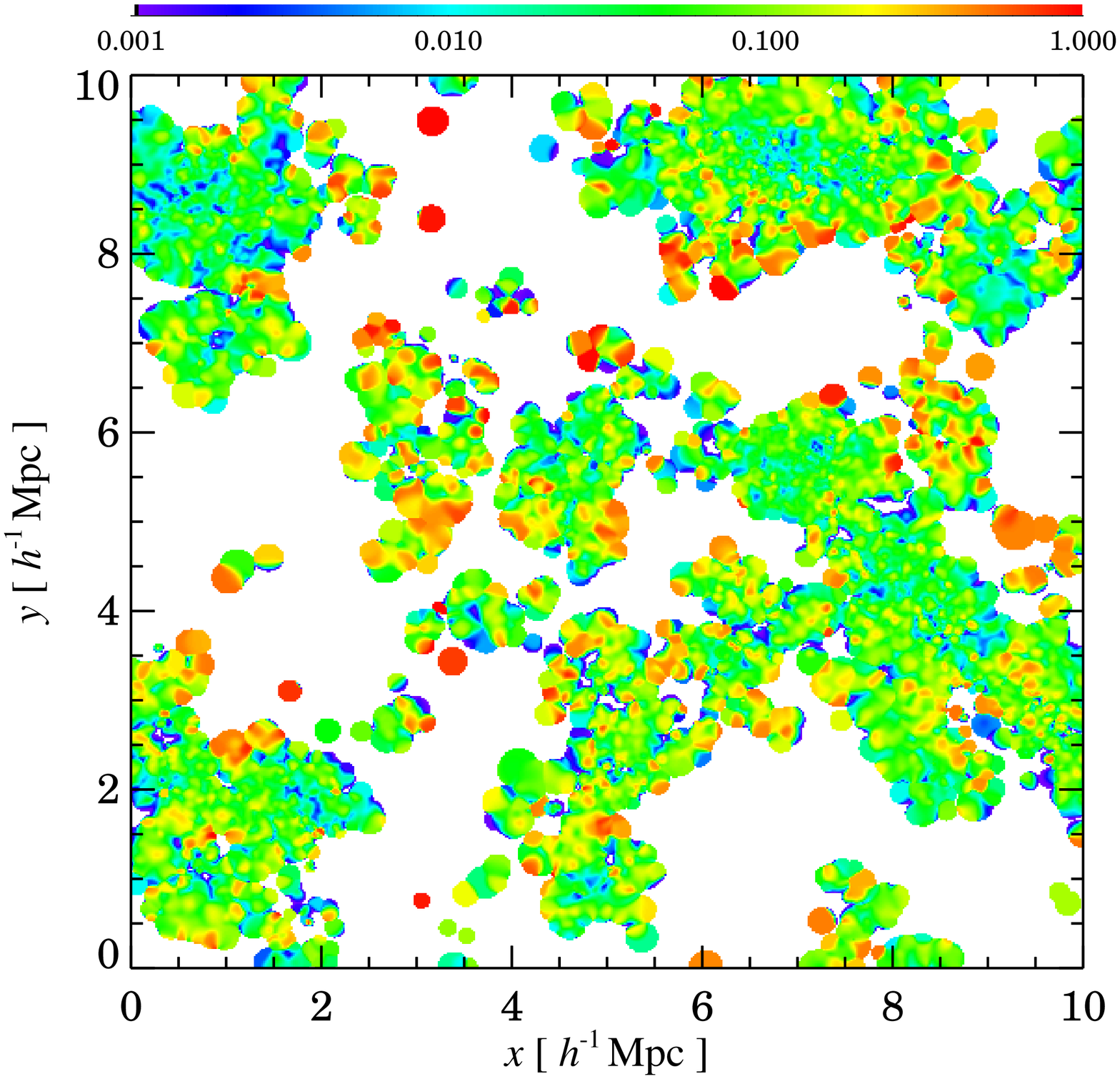,width=8cm,angle=0}}
\caption{Maps resulting from the projection of 500~kpc-thick 
slices through the simulation volume of total mass-averaged metallicity, 
$\langle Z\rangle$ (left panels) and fractional gas metallicity contributed by
Pop III stars, $R_3 = M_{Z,III}/ (M_{Z,III}+M_{Z,II})$ (right) for
three selected redshifts, $z=10, 5, 3$ from top to bottom, respectively.}
\label{metallicity}
\end{figure*}

To make further progress, let us look at the relative spatial
distribution of metals and Pop III star forming sites. In Fig.
\ref{metallicity} we show 500 kpc thick slices through the
simulation volume at three different redshifts $z = 3, 5, 10$. The
maps represent the spatial distribution at these three epochs of
$\langle Z \rangle$ (left panels) and the fractional gas metallicity
contributed by Pop III stars, $R_3 = M_{Z,III}/(M_{Z,III}+M_{Z,II})$;
by construction, $0 < R_3 < 1$. At $z=10$, the volume filling factor
of metals is small, with only a few isolated star-forming regions. The
most active sites are rapidly making the transition to Pop II star
formation mode, as reflected by the fact that $R_3 \ll 1$ within the
largest metal enriched patches; it is only in the smallest and most
recently polluted regions that Pop III enrichment dominates ($R_3
\approx 1$). As evolution proceeds, the metal bubbles tend to grow
around the most ancient star forming sites, propagating in an
inside-out fashion\footnote{ Note that, due to projection effects,
  the bubble interiors are contaminated by foreground Pop III enriched
  pockets.}. The metallicity structure of the bubbles is such that their
interior is dominated by Pop II metals, with Pop III stars confined in
the outermost boundary. Thus, the formation of Pop III stars is forced
to move away from the sites where the first generation of stars
formed. In addition, it becomes less intense as it migrates away from
the density peaks that harbored the first stars. The termination of
the Pop III era occurs when all regions with a (total) density above
the star formation threshold
(assumed to be $n_\star=0.1$~cm$^{-3}$) reach the critical metallicity. 

This evolution reverses the naively expected age-metallicity relation:
at any given redshift there are (almost) metal-free stars that are
{\it younger} than their enriched counterparts. Quantitatively, at
$z=10$ we find that of the enriched (\ie with $Z>0$) gas mass, a
 fraction of $\approx 0.26$ ($\approx 0.26$) is purely polluted by Pop
II (Pop III) stars. These figures change to $\approx 0.46$ ($\approx
0.03$) at $z=5$ and to $\approx 0.7$ ($\approx 0.01$) at $z=3$. At all 
redshifts the remaining enriched gas has a mixed Pop
III/Pop II composition. The overall conclusion is that below $z\approx 9$ 
most of the gas has been enriched only through Pop II supernovae.

\begin{figure*}
\centerline{
\psfig{figure=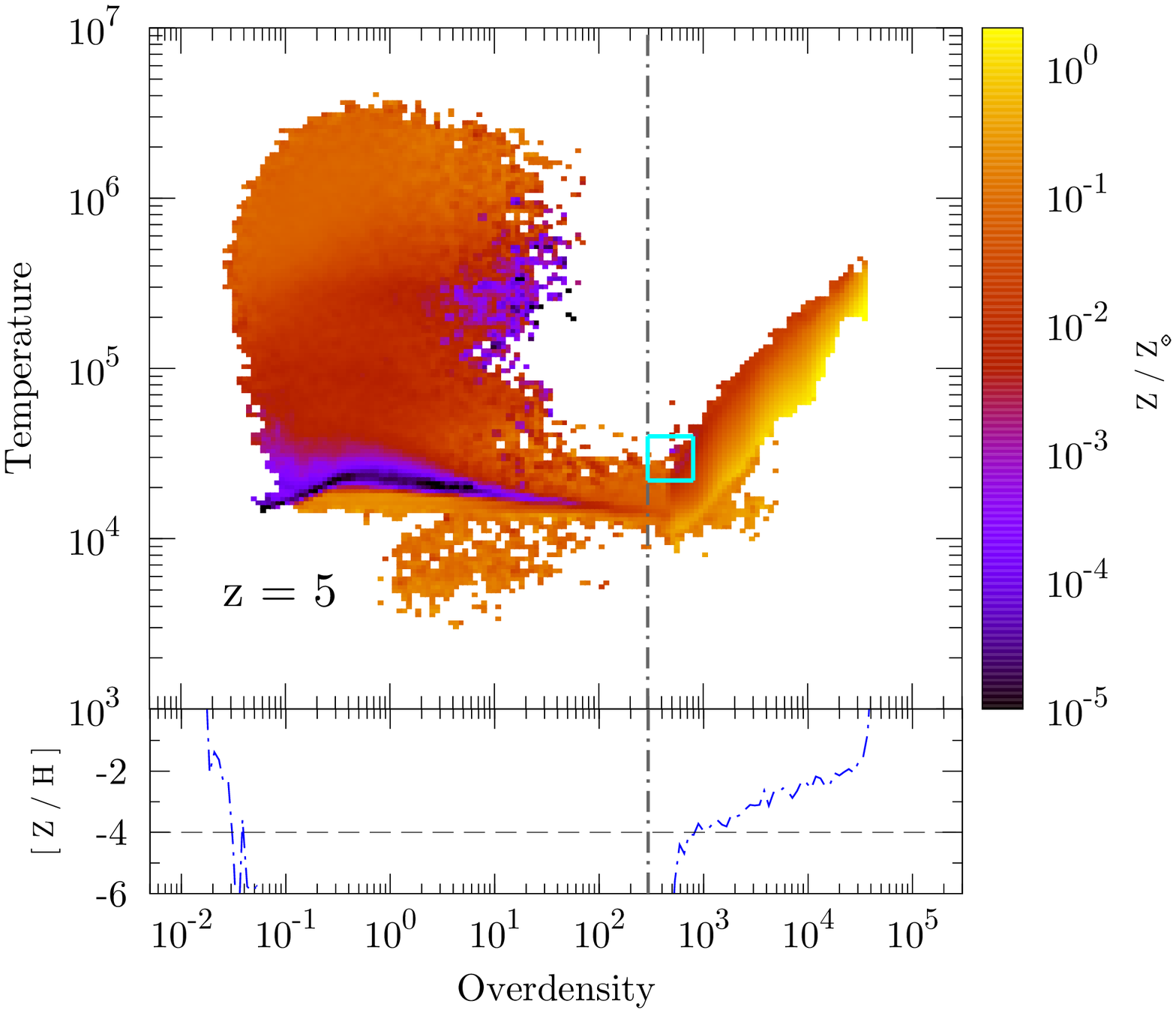,width=8.6cm,angle=0}
$\!\!\!\!\!\!\!\!\!\!\!$
\hspace{1cm}
\psfig{figure=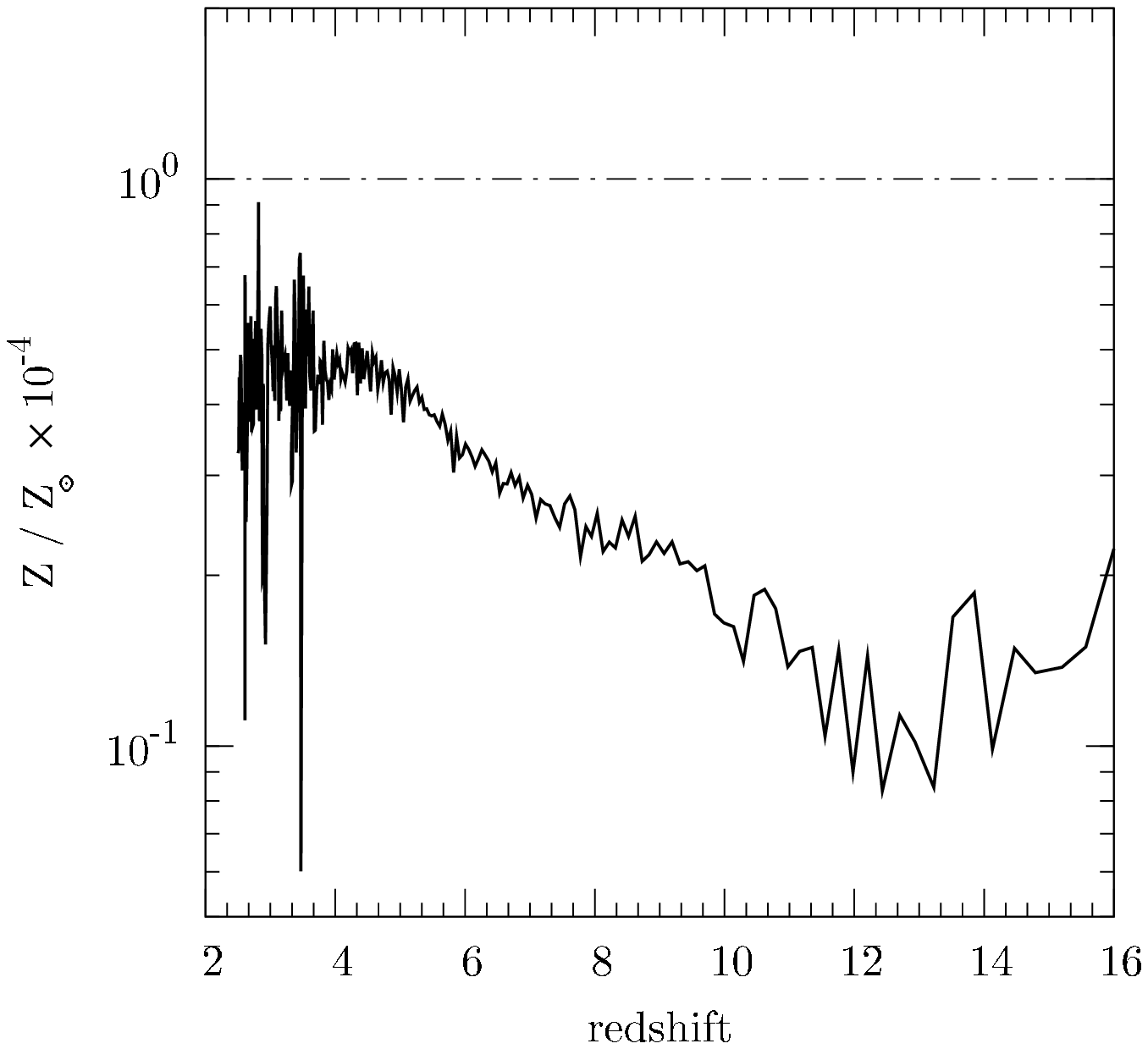,width=8.6cm,angle=0}
}
\caption{ {\it Left}: Mass-averaged metallicity color-coded phase
  diagram of the gas in the simulated volume at $z=5$. Each pixel
  value represents the $\langle Z \rangle$ of the particles within a
  given temperature-overdensity range. The dot-dashed vertical line
  indicates the star formation density threshold $n_\star = 0.1$ cm$^{-3}$. The
  rectangle indicates the area in which
  PopIII-forming particles are found. In the lower panel we plot the
  minimum metallicity found in each overdensity bin (also shown with a 
  dashed line is the value $Z = Z_{\rm cr}$); star forming particles 
  having $Z < Z_{\rm cr}$ are found in the range $300 < \rho_c/\langle\rho \rangle
  <600$ . 
{\it Right}: Mass-averaged mean metallicity of PopIII star forming 
sites as a function of redshift. The dot-dashed line is the value
$Z = Z_{\rm cr}$. }
\label{phase}
\end{figure*}

Fig. \ref{phase} presents the mass-averaged metallicity color-coded
phase diagram of the gas in the simulated volume. Each pixel value
represents the mean temperature and $\langle Z \rangle$ of the
particles within a given mass-overdensity range. There we recognize
the arch-shaped feature ($T\approx 10^4$~K) characterizing low density
IGM (i.e. the Ly$\alpha$ forest) whose thermal budget results from the
balance between photo-heating and adiabatic cooling. The hot $T >
10^{4.5}$~K, tenuous (overdensity $\Delta = \rho_c/\langle \rho \rangle< 10$) 
phase has two distinct
origins as indicated by its metallicity: (i) gas shocked by galactic
outflows ($Z > 10^{-4} Z_\odot$), and (ii) gas virializing in forming
galaxies ($Z < 10^{-4} Z_\odot$): galaxies forming out of this gas can
potentially form Pop III stars. The most relevant region of the phase
plane for this study is the oblique branch found at overdensities 
$\Delta \gsim 3\times 10^2$. Most of this dense gas has $\langle Z
\rangle \gg Z_{\rm cr}$ and is located in active regions of Pop II star
formation. However, a minor fraction of the gas has $\langle Z \rangle
< Z_{\rm cr}$ (region inside the rectangle in Fig.~3) and it is forming Pop III
stars. The Pop III forming sites have densities only slightly above the
critical threshold ($\Delta \approx 300$ at $z=5$)
and cool temperatures. As already pointed out, Pop III star formation
is progressively confined into low density gas, i.e. the periphery of
collapsing structures. The right panel of Fig. \ref{phase} further 
illustrates this point. Almost independently of redshift, Pop III stars tend
to form on average in regions where $\langle Z \rangle \approx 0.1
Z_{\rm cr}$, although a considerable spread around such value is found.

\section{Conclusions}

The main finding of this study is that Pop III star formation
can continue down to very low $z \approx 2.5$ thanks to the
fact that, due to inefficient metal enrichment, pockets of almost
pristine ($Z < Z_{\rm cr}$) gas continue to exist. This confirms
the results of previous semi-analytical models (SSF; Schneider et al.
2006). A general evolutionary picture emerges in which Pop III star
formation starts in the highest density peaks of the cosmic density
field which are then polluted by their metals and turn into Pop II
sites, hence forcing Pop III star formation to migrate towards the
outer, low-density environments. The inside-out propagating
``Pop III-wave'' stops when the environmental density has dropped
below the star formation threshold, $n_\star$. Such segregation in 
regions of density slightly above $n_\star$ causes Pop III star 
formation efficiency to remain low, due to the long free-fall 
and cooling timescales. This is at odd with semi-analytic models
which generally assume a much higher (0.01-0.1) conversion efficiency 
of gas to Pop III stars. A more detailed comparison between numerical
and semi-analytical approaches is deferred to future work.
Pop III stars preferentially form in regions where the mass-averaged
gas metallicity is $10^{-5} Z_\odot$, i.e. 10\% of the critical
metallicity value adopted here. Finally, we have checked the robustness
of all these results against the uncertainties in the modeling of
metal transport and diffusion, adopting different numerical schemes
to describe these processes, as discussed in Tornatore et al. (2007).

The above findings are quite encouraging for searches of Pop III stars
at moderate redshifts as it appears that rather than disappeared, these
stars are hidden in the outskirts of collapsing structures.  Thus, a
non-negligible fraction of observable $z>3$ objects may be powered by
the radiative (Lyman-$\alpha$ emitters, Lyman Break Galaxies: SSF, Malhotra \& Rhoads 2002;
Dawson et al. 2004; Jimenez \& Haiman 2006) or mechanical (pair-instability supernovae: 
Scannapieco et al. 2005) input of Pop III stars. On the other hand,
identifying nucleosynthetic signatures of Pop III stars appears 
extremely challenging, as only a tiny fraction ($8 \times 10^{-5}$) of 
the baryons at $z =3$ has been contaminated purely by Pop III metals.  
Alternative strategies based on the
detection of extremely metal-poor stars in the Galactic halo face a 
similar difficulty in spotting truly second-generation stars
(Salvadori et al. 2007). Thus, it is likely that robust identifications
of metal-free stars at high redshift would be obtained for objects
characterized by large ($>10^3$~\AA) Ly-$\alpha$ equivalent widths 
and/or strong HeII 1640~\AA~line. However, constraining the mass
range, or even the IMF, of such stars on this basis is very difficult.

A final question remains on which physical mechanism dominates the
chemical feedback: transport of metals by outflows or a ``genetic"
form, \ie inheriting metals from the parent sub-halos. From the
discussion above, it is clear that some fraction of the IGM is
polluted by winds to high metallicity preventing it from subsequently
forming Pop IIIs. At present, the role of the ``genetic'' transmission of metals
can not be straightforwardly deduced from our study, as it requires a detailed
investigation of the merging history and metallicity evolution of host halos
which we defer to future work.

\section*{Acknowledgments}
We are grateful to the referee, Naoki Yoshida, for his careful revision and
fruitful comments. We acknowledge and
DAVID\footnote{www.arcetri.astro.it/science/\~cosmology} members for
enlightening discussions. LT is grateful to Stefano Borgani for
insightful suggestions.

\end{document}